\documentclass[amsmath,amssymb,aps,10pt,prd,letterpaper,nofootinbib,balancelastpage,notitlepage,superscriptaddress,twocolumn,floatfix]{revtex4-2}
\usepackage{graphicx}	
\usepackage{ragged2e}	
\usepackage{bm}		    
\usepackage[sort&compress]{natbib}	 	
	\setcitestyle{square,numbers,comma}	
\usepackage[colorlinks=true,urlcolor=blue,linkcolor=black,citecolor=red]{hyperref}	
\renewcommand{\eqref}[2][]{Eq{#1}.~(\ref{eq:#2})}		
\newcommand{\orcid}[1]{\href{https://orcid.org/#1}{\,\includegraphics[width=8px]{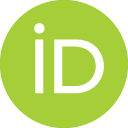}}}

\newcommand{\mc}[1]{\ensuremath\mathcal{#1}}

\newcommand{\eqb}{\begin{equation}}
\newcommand{\eqe}{\end{equation}}
\newcommand{\eqnb}{\begin{eqnarray}}
\newcommand{\eqne}{\end{eqnarray}}
\usepackage{color}
\definecolor{r}{RGB}{255,0,0}
\definecolor{R}{RGB}{255,0,0}
\definecolor{o}{RGB}{255,94,1}
\definecolor{O}{RGB}{255,165,0}
\definecolor{p}{RGB}{223,0,225}
\definecolor{P}{RGB}{223,0,225}
\definecolor{g}{RGB}{0,150,85}
\definecolor{G}{RGB}{0,150,85}
\definecolor{b}{RGB}{0,0,255}
\definecolor{B}{RGB}{0,0,255}
\usepackage{ezedits}

\newcommand{\be}{\begin{equation}}
\newcommand{\ee}{\end{equation}}
\newcommand{\beq}{\begin{eqnarray}}
\newcommand{\eeq}{\end{eqnarray}}

\def\ga{g_{a\gamma\gamma}}
\def\Gammaa{\Gamma_{a\rightarrow2\gamma}}
\def\gammaa{\gamma_{a\rightarrow2\gamma}}
\def\k{\vec k}
\def\kk{\vec k'}
\def\p{\vec p}
\def\e{\epsilon_{\vec k}}
\def\s{s_{\vec k}}
\def\sig{\sigma_{\vec k}}
\def\eps{\epsilon_{\vec k}}
\def\et{\eta_{\vec\kappa}}
\def\mp{m_\phi}
\def\mc{m_\chi}

\begin{document}

\title{Production of massive bosons from the decay of a massless particle beam}
\date{\today}
\author{Ariel Arza\orcid{0000-0002-2254-7408}}
\email{ariel.arza@gmail.com}
\affiliation{Institute for Theoretical and Mathematical Physics, Lomonosov Moscow State University (ITMP), 119991 Moscow, Russia}


\begin{abstract}
Taking a two interacting scalar toy model with interaction term $g\phi\chi^2$, we study the production of $\chi$ particles coming from the decay of an
asymptotic and highly occupied beam of $\phi$ particles. We perform a nonperturbative analysis coming from parametric resonant instabilities and investigate
the possibility that massive $\chi$ particles are produced from decays of massless $\phi$ particles from the beam. Although this process is not present in a
perturbative analysis, our nonperturbative approach allows it to happen under certain conditions. For a momentum $p$ of the beam particles and a mass $m_\chi$
of the produced ones, we find that the decay is allowed if the energy density of the beam exceeds the instability threshold $p^2\mc^4/(2g^2)$.
We also provide an analytical expression for the spontaneous decay rate at the earliest time.
\end{abstract}

\maketitle

\section{Introduction}

The decay of a particle into other species is one of the simplest and most relevant effects in relativistic field theories. From the theoretical point of view, the 
decay rate for a process $\phi_i\rightarrow\phi_j+\phi_k$ can be defined in general terms as \cite{schwartz}
\begin{equation}
\Gamma=V\int{d^3p_j\over(2\pi)^3}V\int{d^3p_k\over(2\pi)^3}{d|S_{fi}|^2\over dt} \enspace, \label{Gamma1}
\end{equation}
where $\vec p_{j,k}$ are the three-momentum of $\phi_{j,k}$, $V$ the volume where the theory is defined, $t$ the time, and $|S_{fi}|^2$ the transition 
probability of the process. When the initial and final states are asymptotic, i.e., free of interactions, the matrix element $S_{fi}\equiv\left<f\right|S\left|i\right>$ is calculated perturbatively. It gives
\be
S_{fi}={i(2\pi)^4\delta^4(p_i-p_j-p_k)\over\sqrt{2\omega_iV}\sqrt{2\omega_jV}\sqrt{2\omega_kV}}{\cal M}_{fi} \label{Smat1}
\ee
where ${\cal M}_{fi}$ is the Feynman amplitude and $\omega_{i,j,k}$ the energy of the particles. In this formula, the delta function imposes that the momentum and energy conservation between
initial and final asymptotic states must be exactly fulfilled in the transition.

In a physical situation, the decay processes originate from beams or clumps of the decaying particles. To compute the produced particle flux the analysis is usually done for the decay of a single
particle of the beam or clump and the decay rate is calculated in perturbation theory using Eq. (\ref{Smat1}), where asymptotic states of the produced particles is assumed. Finally it is integrated
over the contributions of all the particles of the beam. This reasoning is valid as long as the assumption of asymptotic final states is correct or at least when it is a good approximation. Following
this recipe, it is easy to find that massless particles of a beam can not decay into other massive species because the energy-momentum conservation for asymptotic states is never
fulfilled\footnote{Under the aymptotic state assumption, the case of massless particles decaying into other massless particles has been studied in detail in \cite{Parkinson:1973iy,Vasholz:1974ws,Parkinson:1974tk,Fiore:1995ai}.}. More general,
this recipe based on the asymptotic state assumption forbids any decay process for which the total mass of the produced particles exceeds the mass of the decaying particles.

However, if the beam or clump carries an extremely big energy density, this assumption is no longer valid, the produced particles are not in asymptotic states because they continue to interact
with the decaying field and then nonperturbative effects start to be relevant. As a consequence, Eq. (\ref{Smat1}) is meaningless and the energy-momentum conservation as well as the
decay rate should be accounted in a different way. This motivates us to explore the possibility that processes (not only decays) forbidden by the asymptotic final state assumption are now allowed.

In this article we study these nonperturbative effects for a simple scalar toy model. We focus on a beam of massless particles and their decay into massive species. We find that this is indeed possible
when the beam energy density exceeds some threshold which is explicitly calculated. As far as we know this possibility has not been discussed in the literature and, despite our study being limited to a
theoretical point of view, we expect that it could be relevant in astrophysics, collider physics, and cosmology since in these topics, it is usual to ignore processes that, under the asymptotic state assumption,
seem to be impossible. As very intense fields play an important role in these subjects, we suggest that these effects should be something to look at.

\section{Non-perturbative axion decay as an example} \label{axioncase}

To start, let us introduce, as an example, the interesting case of the axion decay into two photons. This inspired us to write this article and certainly serves as a good example to understand what we
mean with nonperturbative effects in the previous paragraphs. Assuming asymptotic photon states for the decay of a single axion, the scattering matrix of the process can be computed perturbatively from
Eq. (\ref{Smat1}), then the decay rate obtained from Eq. (\ref{Gamma1}) is given by
\be
\Gammaa={\ga^2m_a^3\over64\pi} \label{Gamma-axion-photon1}
\ee
in the axion rest frame. Here $\ga$ is the axion to two photons coupling and $m_a$ the axion mass. For an axion clump composed by $N_a$ particles (all of them at rest), and
assuming, again, asymptotically free photon states after decays, the number of photons per unit time emitted by the clump is $2N_a\Gammaa$. This straightforward result ignores the fact that once photons are
produced, they continue to interact with the clump. Indeed, if the axion clump is highly occupied in a particular mode, the presence of the produced photons stimulate the decay of
the remaining axions, leading to an exponential growth of the photon occupancy number \cite{Yoshida:2017ehj,Hertzberg:2018zte,Arza:2018dcy,Carenza:2019vzg,Wang:2020zur,Arza:2020eik,Levkov:2020txo}.
This process originates from Bose statistics and is known as Bose enhancement. The effect can also be interpreted as a parametric resonance on the electromagnetic field.
Parametric resonance is a well known effect with many applications in cosmology, for instance in the physics of inflation
\cite{Traschen:1990sw,Kofman:1994rk,Shtanov:1994ce,Khlebnikov:1996mc,Khlebnikov:1996wr,Kofman:1997yn} (see Ref. \cite{Amin:2014eta} for a review).

Let us discuss briefly this parametric resonance in more detail
for the axion-photon case. As it is a nonperturbative calculation, it is better understood for time scales where the clump has not been significantly depleted. Under this assumption
one can neglect backreactions on the axion field and the resulting differential equations for the electromagnetic field can be linearized. For a dense axion clump with energy density
$\rho_a$ pairs of photons emitted with momenta $\k$ and $-\k$ have a time dependent occupancy number (in the linear regime) given by
\be
f_{\gamma,k}(t) = {\sigma_a^2\over s_k^2}\sinh(s_kt)^2 \enspace.       \label{photon-number}
\ee
Here $\sigma_a=\ga/2\sqrt{\rho_a/2}$, $\eps=2k-m_a$ and $s_k=\sqrt{\sigma_a^2-\eps^2/4}$, where $k=|\vec k|$. The most important feature of this result is that each mode
$\vec k$ that satisfies $\eps^2<4\sigma_a^2$, i.e., 
\be
{m_a\over2}-\sigma_a<k<{m_a\over2}+\sigma_a \label{res-cond}
\ee
experiences exponential growth in its occupancy number. We can compute the photon number density using a saddle point approximation, for $\sigma_at\gg1$ we find
\be
n_{\gamma}(t)=2\int{d^3k\over(2\pi)^3}f_{\gamma,k}(t)\simeq {m_a^2\sigma_a\over8\sqrt{(2\pi)^3}}e^{2\sigma_at} \enspace. \label{photon-density}
\ee
From Eqs. (\ref{photon-density}) and (\ref{res-cond})  we see that the photon number density grows at the rate
\be
\gammaa=2\sigma_a\sim \ga\sqrt{\rho_a\over2} \label{Gamma-axion-photon2}
\ee
affecting a bandwidth of size
\be
\delta k=2\sigma_a \label{axionband}
\ee
centered at $k=m_a/2$.

This non perturbative axion decay features two interesting properties that we can not find in the perturbative case. First, notice that from Eq. (\ref{Gamma-axion-photon2})
we see that the photons are emitted at a rate different from what we observe in Eq. (\ref{Gamma-axion-photon1}). The most important discrepancy is that while the perturbative
result scales as $\ga^2$, the nonperturbative one scales as $\ga$. References \cite{baumann,Alonso-Alvarez:2019ssa} give a very intuitive explanation about what is happening.
Looking at the Boltzmann equation, one finds 
\be
\dot n_a=-\Gammaa\left(1+2f_{\gamma,m_a/2}\right)n_a \enspace. \label{Boltzmann-axion}
\ee
We can see that the decay rate is corrected by a factor $1+2f_{\gamma,m_a/2}$. When $f_{\gamma,m_a/2}>1/2$, it is enhanced dramatically. This stimulated decay effect is
known as Bose enhancement. To probe that $f_{\gamma,m_a/2}>1/2$ can be reached easily, let us analyze Eq. (\ref{Boltzmann-axion})
at the beginning, when $f_{\gamma,m_a/2}\sim0$. Since $\dot n_\gamma=-2\dot n_a$, after a small period of time $\delta t$ we get $n_\gamma=2\Gammaa n_a\delta t$. From the uncertainty
principle the photon field is spread in the energy bandwidth $\delta k\sim1/\delta t$, then
\beq
f_{\gamma,m_a/2} &\sim& {n_\gamma\over8\pi(m_a/2)^2\delta k/(2\pi)^3} \nonumber
\\
&=& 8\pi^2{\Gammaa n_a\over m_a^2\delta k^2} \enspace. \label{fgamma}
\eeq
Now, using Eq. (\ref{Gamma-axion-photon1}) and taking Eq. (\ref{axionband}) for $\delta k$, we find
\be
f_{\gamma,m_a/2}\sim{\pi\over4}>{1\over2} \enspace.
\ee
It means that the effect coming from Bose enhancement becomes important instantly. Moreover, using Eq. (\ref{fgamma}), Eq. (\ref{Boltzmann-axion}) can be
written as
\be
\dot n_\gamma\sim2\sigma_a n_\gamma \enspace,
\ee
which is consistent with Eq. (\ref{photon-density}), at least in the order of magnitude.  The other property, maybe the most important concerning our hypothesis, comes from Eq. (\ref{res-cond}).
While in the perturbative calculation the energy momentum conservation for asymptotic states implies that all the produced photons must have a momentum $k=m_a/2$, the nonperturbative
analysis allows the produced photon to have momenta within the bandwidth defined in Eq. (\ref{res-cond}). As this window for the photon momenta is proportional to $\sqrt{\rho_a}$, we would
recover the perturbative result in the limit $\rho_a\rightarrow0$. This limit is indeed consistent with the asymptotic state assumption since for a small axion energy density the interaction
of the produced photons with the axion field is negligible. This window for the momenta of the produced particles, photons for the axion case, is what leads us to wonder about the possibility
of decays that are forbidden in the perturbative calculation, where the asymptotic state assumption is an unbroken rule.

For a more general decay process of the form $\phi_i\rightarrow\phi_j+\phi_k$, if the decaying field $\phi_i$ is highly occupied, the linearized equations of motion for the fields $\phi_j$
and $\phi_k$ look quite similar to the ones of the axion-photon case. While the analog of $\sigma_a$ has in general a momentum dependence ($\sigma_a\rightarrow\sigma_{\k}$) and a different
form (it also depends on the form of the interaction term), it always scales as the squared root of the energy density $\rho_{\phi_i}$ of the decaying field $\phi_i$. On the other hand, and also very
important, $\epsilon_k$ has always the same form; i.e., if the decaying
field $\phi_i$ is occupied in a state with momentum $\vec p$, 
\be
\epsilon_{\vec k}=\omega_{\vec k}^{(j)}+\omega_{\vec p-\vec k}^{(k)}-\omega_{\vec p}^{(i)} \enspace, \label{eps-gen1}
\ee
where $\omega_{\vec k}^{(l)}=\sqrt{k^2+m_l^2}$ ($l=\{i,j,k\}$) and $m_l$ the corresponding mass of each field $\phi_l$. The parametric
resonance and depletion of the field $\phi_i$ takes place when 
\be
\epsilon_{\vec k}^2<4\sigma_{\vec k}^2 \enspace. \label{cond-gen1}
\ee
Notice that in the limit $\rho_{\phi_i}\rightarrow0$, suitable for the single particle decay, the condition $\eps=0$ is required. This condition is exactly the energy-momentum conservation for asymptotic states.

We are interested in the case where $\phi_i$ is massless\footnote{In models of massless preheating \cite{Moghaddam:2014ksa} the inflaton field suffers self-interactions due to a $\lambda\phi^4$ potential.
It makes the zero momentum modes of the inflaton to oscillate harmonically. It finally drives the decay of the field into SM particles by parametric resonance. Our interest is rather directed at a beam of
massless particles and their possibility to decay into massive species.} and $\phi_j$, $\phi_k$ are massive (or at least one of them). In this case $\epsilon_{\vec k}$ can never be zero and, therefore, a single
massless particle can not decay by itself. However, when $\phi_i$ is a highly populated beam in some momentum mode, the condition for the decay is actually Eq. (\ref{cond-gen1}). Particles from the massless beam
start to decay into $\phi_j$ and $\phi_k$ with momentum modes satisfying Eq. (\ref{cond-gen1}) even if $\epsilon_{\vec k}=0$ has no solutions for $\k$.

When non-perturbative effects take place, of course the energy conservation can not be accounted by assuming asymptotic states. In our approach we do it by calculating the energy expectation value of the
produced field directly from the solutions of the equations of motion. The details will be discussed later. 

At this point we have taken the parametric resonance properties to explain roughly how a beam of massless particles can be depleted by the decay into massive particles. In the following
we will discuss this mechanism in detail using a simple toy model involving two scalar fields. Our analysis includes solutions of the equations of motion in the rotating wave approximation,
study of the parametric resonance instabilities, computation of the energy density threshold for the decays and consistency check of energy conservation. Some technical details and validity
check of the approximation are left in the Appendix.

\section{A two scalar toy model}

We consider the following interaction hamiltonian density
\be
{\cal H}_I=g\phi\chi^2 \label{hamiltonian-density}
\ee
where $\phi$ represents the decaying particles, $\chi$ the produced particles and $g$ the coupling constant. For now we consider the general case where both fields are massive.
Working in the Heisenberg picture, the equations of motion of the system are
\beq
(\Box+\mp^2)\phi &=& -g\chi^2 \label{eqphi1}
\\
(\Box+\mc^2)\chi &=& -2g\phi\chi \label{eqchi1}
\eeq
where $\mp$ and $\mc$ are the masses of $\phi$ and $\chi$, respectively. We are interested in the case where $\phi$ is a beam of particles, highly occupied in a single momentum
mode $\p$. In this limit, $\phi$ can be perfectly considered as a classical field. On the other hand, as in the axion-photon case, we consider timescales where $\phi$ is not
significantly depleted. It allows us to neglect backreactions over $\phi$, so we make the rhs of Eq. (\ref{eqphi1}) equal to $0$ and Eq. (\ref{eqchi1}) becomes linear.
These considerations lead us to write $\phi$ as a monochromatic classical plane wave as
\be
\phi(\vec x,t)={\sqrt{2\rho_\phi}\over\omega_p}\cos(\p\cdot\vec x-\omega_{\p}\,t) \label{phi} \enspace,
\ee
where $\rho_\phi$ is the time averaged energy density of the beam and $\omega_{\p}=\sqrt{p^2+\mp^2}$.

We write the quantum field $\chi$ in terms of creation and annihilation operators as
\be
\chi(\vec x,t)=\int{d^3k\over(2\pi)^3}{1\over\sqrt{2\Omega_{\vec k}}}\left(\chi_{\vec k}(t)\,e^{i\vec k\cdot\vec x}+\chi_{\vec k}(t)^\dagger\,e^{-i\vec k\cdot\vec x}\right)  \label{chi1}
\ee
where $\Omega_{\vec k}=\sqrt{k^2+m_\chi^2}$ and the operators $\chi_{\k}$ and $\chi_{\k}^\dagger$ satisfy the
commutation relations $[\chi_{\k},\chi_{\kk}]=0$ and $[\chi_{\k},\chi_{\kk}^\dagger]=(2\pi)^3\delta^3(\k-\kk)$.
Inserting Eqs. (\ref{phi}) and (\ref{chi1}) into Eq. (\ref{eqchi1}), we get the following set of equations for the new operators $A_{\k}=\chi_{\k}+\chi_{-\k}^\dagger$:
\beq
\left(\partial_t^2+\Omega_{\k}^2\right)A_{\k} &=& -\omega_{\p}^2\,\alpha\left(\sqrt{\Omega_{\k}\over\Omega_{\k-\p}}A_{\k-\p}\,e^{-i\omega_{\p}\,t}\right. \nonumber
\\
& & \left.+\sqrt{\Omega_{\k}\over\Omega_{\k+\p}}A_{\k+\p}\,e^{i\omega_{\p}\,t}\right), \label{eqA1}
\eeq
where we have defined
\be
\alpha\equiv{g\sqrt{2\rho_\phi}\over\omega_{\p}^3} \enspace. \label{alpha}
\ee

Let us study the resonant solutions of Eq. (\ref{eqA1}) analytically using the rotating wave approximation (RWA)\footnote{Figure \ref{fig} (see the Appendix) shows a perfect agreement between numerical solutions of Eq. (\ref{eqA1}) and solutions in the rotating wave approximation.}. To do so we write $\chi_{\k}(t)=a_{\k}(t)\,e^{-i\Omega_{\k}\,t}$.
The RWA is a method that allows to transform a second order system of differential equations to a first order one, in our context it can be done when $a_{\k}$ varies slowly respect to $\chi_{\k}$. For more details of the method, see Appendix \ref{RWA}.

We are interested in the case where at $t=0$ there is no $\chi$ particles. Therefore, at the very beginning the only relevant process\footnote{The decay $\phi\rightarrow2\chi$ is the main process only at the beginning. When $\chi$ is populated, processes of the form $\phi+\chi\rightarrow\chi$ are activated.
See Appendix \ref{RWA} and \ref{processes} for a more detailed discussion.} to account is
$\phi\rightarrow2\chi$.
Defining 
\be
\epsilon_{\k}=\Omega_{\vec k}+\Omega_{\vec p-\vec k}-\omega_{\vec p} \label{eps1}
\ee
and neglecting non-relevant terms, Eq. (\ref{eqA1}) becomes
\be
\dot a_{\k}=-i\sig\,a_{\p-\k}^\dagger\,e^{i\epsilon_{\k}\,t} \label{eqa2}
\ee
where
\be
\sig=g\sqrt{\rho_\phi/2\over\omega_{\p}^2\,\Omega_{\p-\k}\,\Omega_{\k}} \enspace. \label{sigmak}
\ee
The general solution of Eq. (\ref{eqa2}) is
\beq
a_{\k}(t) &=& e^{i\e t/2}\left[a_{\k}(0)\left(\cosh(\s t)-i{\e\over2\s}\sinh(\s t)\right)\right. \nonumber
\\
& & \left.-i{\sig\over\s}a_{\p-\k}(0)^\dagger\sinh(\s t)\right], \label{sol-a1}
\eeq
where we have defined
\be
\s=\sqrt{\sig^2-\eps^2/4} \enspace. \label{s1}
\ee
For each momentum mode $\k$, its occupancy number is
\be
f_{\chi,\k}(t) ={1\over V}\left<0\right|a_{\k}(t)^\dagger a_{\k}(t)\left|0\right>={\sig^2\over \s^2}\sinh(\s t)^2      \label{a-number}
\ee
and the total density of produced particles is given by
\be
n_{\chi}(t)=\int{d^3k\over(2\pi)^3}f_{\chi,\k}(t) \enspace. \label{densitya}
\ee
The modes that exhibit parametric resonance are the ones that satisfy
\be
\s^2>0 \enspace. \label{scond1}
\ee

Now let us talk about energy conservation (momentum conservation is satisfied since we obtained Eq. (\ref{eqA1})). After $N_\phi$ decays of $\phi$ particles, the energy expectation value of $\chi$ must be $\left<E_\chi\right>=\omega_{\p}N_\phi$
in order to conserve energy during the processes. We compute $\left<E_\chi\right>$ directly by $\left<E_\chi\right>=\left<0\right|H_\chi\left|0\right>$,
where the free hamiltonian $H_\chi$ is given by 
\be
H_\chi={1\over2}\int d^3x\left((\partial_t\chi)^2+(\vec\nabla\chi)^2+m_\chi^2\chi^2\right) \enspace. \label{Hchi}
\ee
By using Eqs. (\ref{chi1}) and (\ref{sol-a1}) we get (see Appendix \ref{energy})
\be
\left<E_\chi\right>=V\int {d^3k\over(2\pi)^3}(\Omega_{\k}-\e/2)f_{\chi,\k} \label{Echi1}
\ee
where we have ignored the vacuum energy contribution. Defining $\vec q=\p-\k$ we find
\be
\left<E_\chi\right>=\omega_{\p}N_\phi+{V\over2}\int {d^3k\over(2\pi)^3}\left(\Omega_{\k}\,f_{\chi,\k}-\Omega_{\vec q}\,f_{\chi,\vec q}\right) \enspace, \label{Echi2}
\ee
where we have used Eq. (\ref{eps1}), the property $f_{\chi,\k}=f_{\chi,\vec q}$ and the fact that the average number of produced $\chi$ particles $N_\chi=Vn_\chi$ is twice
$N_\phi$. Since $\int d^3k=\int d^3q$, the integral in Eq. (\ref{Echi2}) cancels, getting
\be
\left<E_\chi\right>=\omega_{\p}N_\phi  \label{Echi3}
\ee
as expected.

Notice that this result as well as the ones in the previous paragraph do not depend on the masses of the particles. The only requirement is a parametric instability. We will show later that
even the case of massless $\phi$ and massive $\chi$ is unstable for some region in momentum parameter space if $\rho_\phi$ is large enough. In other words, we will
show that, after some threshold of $\rho_\phi$, the production of massive scalar from decays of massless particles is possible without energy-momentum violation.

Before continuing, we would like to discuss the validity of our calculation. Our results were found in a RWA
where the amplitudes $a_{\k}$ vary slowly respect to $\chi_{\k}$, so Eq. (\ref{sol-a1}) is valid under the condition
$\Omega_{\k}\gg\s$. Assuming $\Omega_{\k}\sim\Omega_{\p-\k}\sim\omega_{\p}$ we would be safe for
\be
\alpha\ll1 \enspace. \label{break}
\ee
Even though condition Eq. (\ref{break}) will be useful in most of the remaining discussion, we have to pay special attention when either $\Omega_{\k}$ or $\Omega_{\p-\k}$
approaches to zero because $\s$ could blow up. For instance, when $\mc=0$, $\Omega_{\p-\k}=0$ if $\k=\p$.

\section{Instability condition and decay of a massless beam}

Now we are going to study the instability conditions in two opposite cases: (1) massive $\phi$ and massless $\chi$, and (2) massless $\phi$ and massive $\chi$.
Case (1) has been strongly discussed in the literature, so we are using it as a matter of comparison with case (2), which is what concerns us. Let us define the
dimensionless quantities $\vec\kappa=\k/\omega_p$, $\mu=m_\chi/\omega_p$, $\vec v=\vec p/\omega_p$, and $\beta_{\vec\kappa}=\sqrt{\kappa^2+\mu^2}$. Let us
also define $\et=s_{\vec \kappa}/\omega_p$, which in terms of the defined dimensionless parameters is given by
\begin{figure}[t]
   \includegraphics[width=0.5\textwidth]{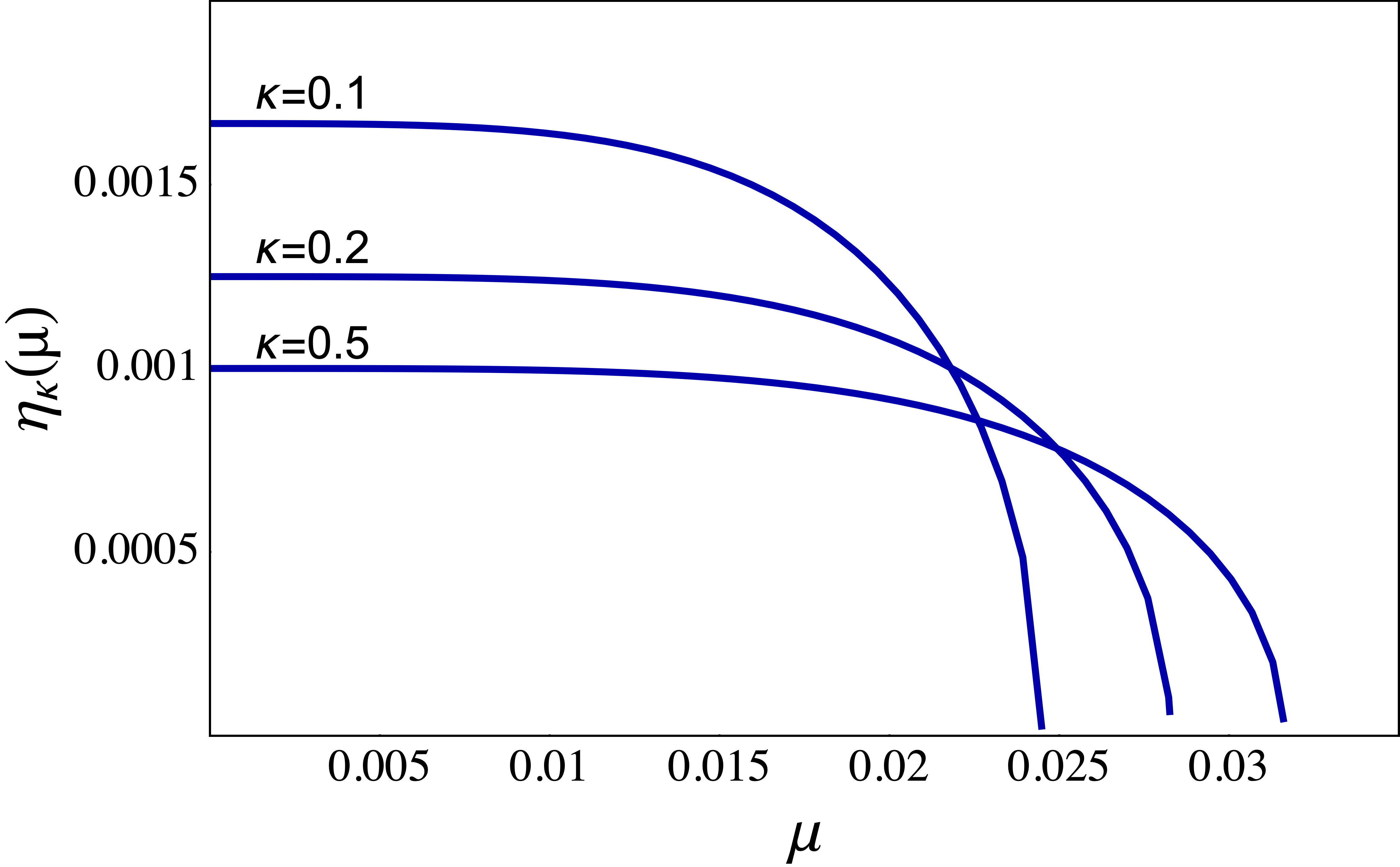}
   \caption{$\eta_{\vec\kappa}$ as a function of $\mu$ for $\kappa=0.1$, $0.2$ and $0.5$. We used $\theta=0$ and $\alpha=10^{-3}$.}
    \label{fig1}
\end{figure}
\be
\et=\sqrt{{\alpha^2\over4\beta_{\vec v-\vec\kappa}\beta_{\vec\kappa}}-{1\over4}(\beta_{\vec\kappa}+\beta_{\vec v-\vec\kappa}-1)^2} \enspace. \label{eta}
\ee
We also define $\theta$ as the angle formed by $\vec k$ and $\vec p$.

\begin{figure}[t]
  \includegraphics[width=0.5\textwidth]{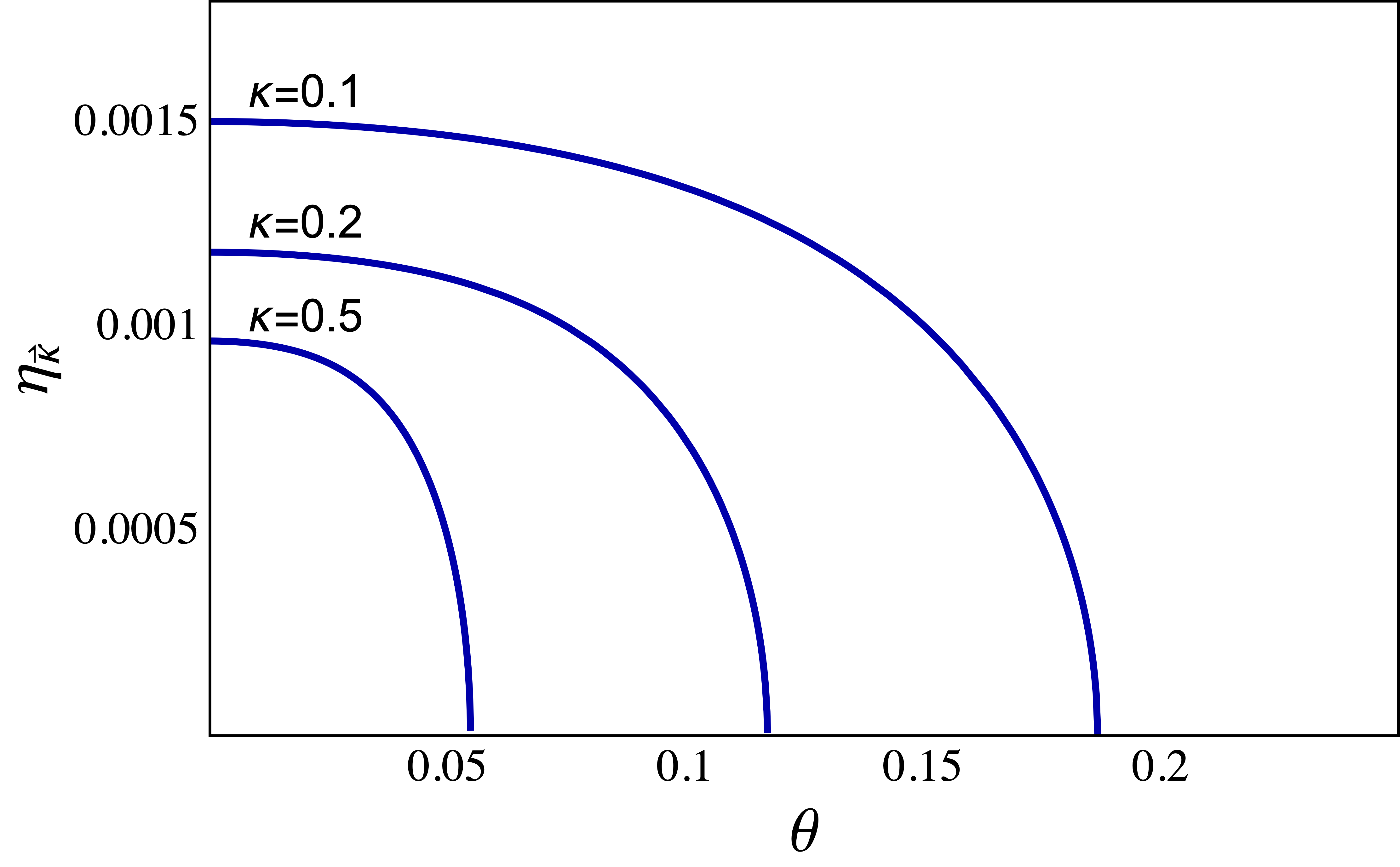}
   \caption{$\eta_{\vec\kappa}$ as a function of $\theta$ for $\kappa=0.1$, $0.2$ and $0.5$. We used $\alpha=10^{-3}$ and $\mu=\sqrt{\alpha}/2$.}
    \label{fig2}
\end{figure}

\begin{figure}[t!]
    \includegraphics[width=0.5\textwidth]{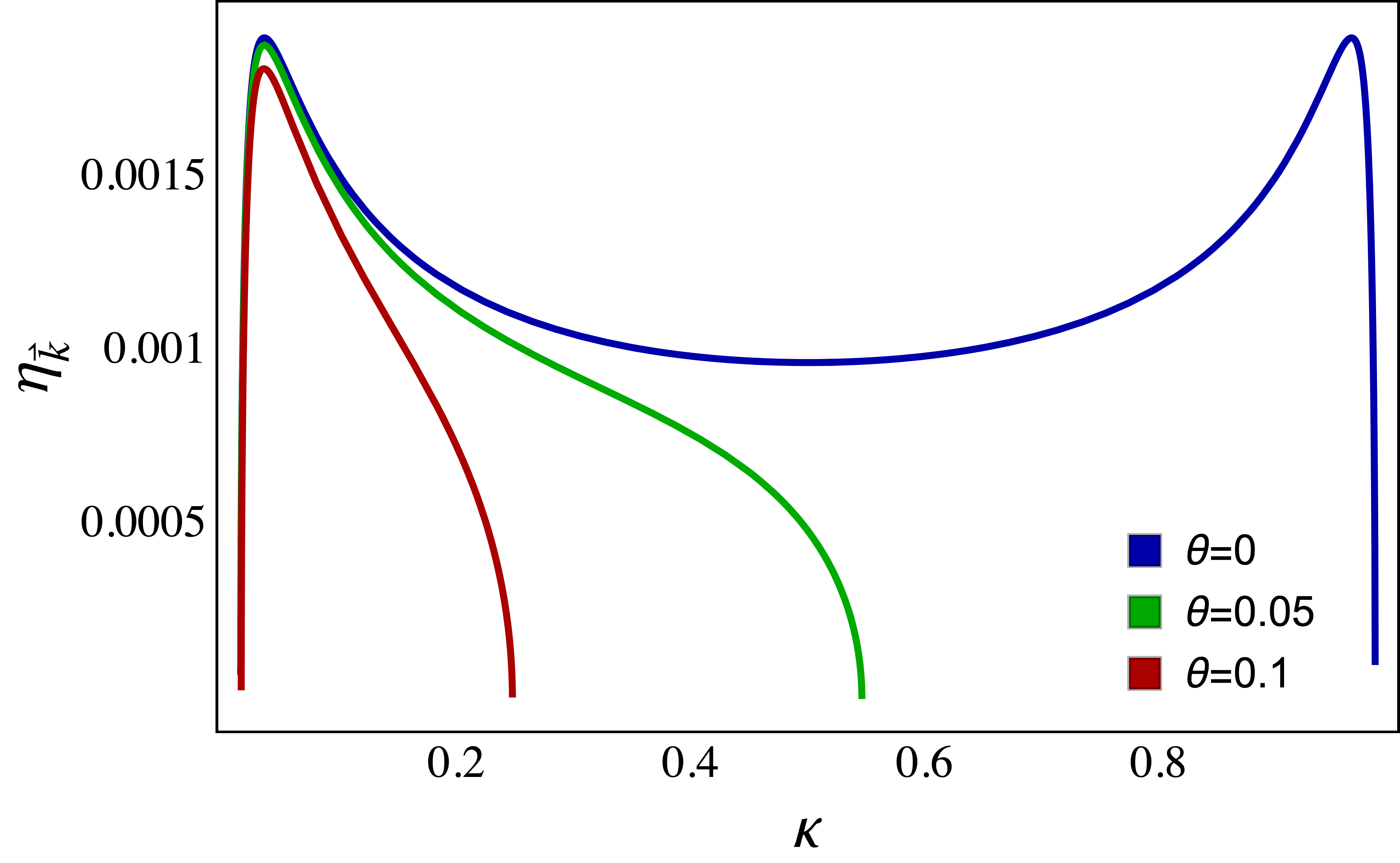}
   \caption{Shape of $\eta_{\vec\kappa}$ as a function of $\kappa$ for $\theta=0$, $0.05$ and $0.1$. Here we have used $\alpha=10^{-3}$ and $\mu=\sqrt{\alpha}/2$.}
    \label{fig3}
\end{figure}

For case (1) we can always choose a reference frame where the $\phi$ particles are at rest, then the beam becomes a clump of particles with speed $v=0$.
Notice that this is very similar to the axion-photon case discussed before\footnote{It is also analogous to the inflaton decay where the inflaton field is considered as an homogeneous classical field \cite{baumann}.}.
Eq. (\ref{eta}) simplifies and the instabilities ($\eta_{\vec\kappa}^2>0$) take place for $1/2-\alpha<\kappa<1/2+\alpha$.
For case (2) Eq. (\ref{eta}) must be evaluated with $v=1$. In this case the analysis is more complicated, so before finding analytical expressions we will first identify
some general properties by showing some plots.  Figure \ref{fig1} shows $\et$ as a function of $\mu$ for $\alpha=10^{-3}$, $\theta=0$, and different values of $\kappa$.
Of course we avoid the extremes $\kappa=1$ and $\kappa=0$ where our calculation breaks down. We can see instabilities only for $\mu\ll1$. Figure \ref{fig2} shows $\et$ as a function
of $\theta$ for $\alpha=10^{-3}$, $\mu=\sqrt{\alpha}/2$, and different values of $\kappa$. We observe that the instabilities occur for $\theta$ smaller than $1$. To have a better
look of the instability regions, see
Fig. \ref{fig3} where $\et$ is plotted as a function of $\kappa$ for a fixed $\mu$ and different values of $\theta$. We see that in terms of $\kappa$ the bandwidth is of order $1$ for $\theta=0$,
much bigger than case (1) where it is of the order of $\alpha$. However, we also see that the instability region becomes smaller as $\theta$ increases, which is the main advantage of case (1)
where the instability window keeps its size for any angle. We also see in Fig. \ref{fig3} that, close to the extremes, $\kappa\approx1$ and $\kappa\approx0$, $\et$ is enhanced. It may contribute
to the efficiency of the process, although we have to be always alert that $\s/\Omega_{\k}\ll1$ is satisfied, otherwise we would enter in regions where our RWA breaks down.

These plots gave us some generals features of the instability conditions, in terms of physical parameters we found that $\mc\ll\omega_{\p}$ and $\theta\ll1$.
Now we will use these features to find analytical expressions. The frequencies can be approximated as $\Omega_{\k}\approx k+\mc^2/(2k)$ and
$\Omega_{\p-\k}\approx p-k+(pk\theta^2+\mc^2)/(2(p-k))$. When computing $\eps$, the main contributions of $\Omega_{\k}$ and $\Omega_{\p-\k}$ cancel with $\omega_{\p}$, getting
\be
\eps\approx{p\,(\mc^2+k^2\theta^2)\over2k(p-k)} \enspace. \label{epsapprox}
\ee
In the above result we have assumed that we are far enough from $k=0$ and $k=p$.

To find the instability condition we solve the equation $\eta_{\vec\kappa}^2=0$ to get the instability limits for $\kappa$. As $\theta=0$ gives the biggest instability
window for $\kappa$, we evaluate at this angle obtaining the limits
\be
\kappa_{\pm}={1\pm\sqrt{1-\mu^4/\alpha^2}\over2} \enspace. \label{kpm}
\ee
The size of the window is
\be
\delta\kappa=\sqrt{1-\mu^4/\alpha^2} \enspace, \label{widthkappa}
\ee
therefore the instability is triggered when $\mu<\sqrt{\alpha}$ or when
\be
\rho_\phi>{p^2\mc^4\over2g^2} \enspace, \label{insta-cond}
\ee
in terms of the energy density.

It is not too difficult to check that the instability condition Eq. (\ref{insta-cond}) is Lorentz invariant. It is, of course, what we would expect. Now two questions arise;
why the instabilities were found
for $p\gg\mc$?. What happens for $p\sim\mc$ or $p\ll\mc$?. We answer the first question as follows. First, the instability occurs when $\mu<\sqrt{\alpha}$.
Second, we are working in a RWA where the validity of our analysis is limited by $\alpha\ll1$. These two facts imply that the instability must satisfy
$p\gg\mc$ as a consequence.
In other words; in our rotating wave approximation, the instability takes place only for reference frames where $p\gg\mc$. To answer the second question, we take Eq. (\ref{eqA1})
in the limit $p\ll\mc$. Taking into account only
the modes $\k$ and $\k-\p$ we get the system
\beq
\left(\partial_t^2+\mc^2\right)A_{\k} &=& -\omega_{\vec p}^2\,\alpha\,A_{\k-\p}\,e^{-i\omega_{\p}\,t} \label{eqAkpsmall1}
\\
\left(\partial_t^2+\mc^2\right)A_{\k-\p} &=& -\omega_{\vec p}^2\,\alpha\,A_{\k}\,e^{i\omega_{\p}\,t} \enspace. \label{eqAkpsmall2}
\eeq
For solutions of the form $A_{\k}=c_{\k}\,e^{\gamma t}\,e^{-i\omega_{\vec p}\,t/2}$ and $A_{\k-\p}=d_{\k}\,e^{\gamma t}\,e^{i\omega_{\vec p}\,t/2}$, we find that
$\gamma$ is given by
\be
\gamma^2=\omega_{\vec p}^2\,\alpha-\mc^2 \enspace. \label{gammapsmall}
\ee
It is clear that the instability is also activated when Eq. (\ref{insta-cond}) is fulfilled. We also expect the same result for $p\sim m_\chi$, but it is not clear how to proceed analytically.

\section{Spontaneous decay rate}

So far we have explained that the decays of massless into massive particles are allowed due to a nonperturbative effect caused when a huge energy density of
the beam prevents asymptotic states for the produced particles. However, we still do not clarify how the first $\chi$ pair is spontaneously released. To do so, we focus
on the decay of a single $\phi$-particle of the beam but, of course, taking into account the effect that the beam energy density causes on it.  Working now in the interaction picture,
the only non-zero element of the $S$-matrix expansion is
\be
S_{fi}=\int_0^Tdt\int d^3x\left<f\right|{\cal H}_{I}\left|i\right>, \label{Smat2}
\ee
where $\left|i\right>$ is the initial state that contains one $\phi$ particle with momentum $\p$ and $\left|f\right>$ the final state composed by two $\chi$ particles, one
with momentum $\k$ and another with momentum $\vec q$. In Eq. (\ref{Smat2}), $T$ is the time during which the system transits from the state $\left|i\right>$
to the state $\left|f\right>$. As the transitions that are typically studied involve processes where the final state is free of interactions, one could just make $T\rightarrow\infty$.
However, in some cases this approximation breaks down due to finite time effects \cite{Joichi:1997xn,Fischbach:2001ie,Giacosa:2010br} or simply if the transition time
is not big enough to be considered as infinite \cite{Erken:2011dz}. In our case, this is indeed what happens. From the nonperturbative analysis we found that if the
instability condition is fulfilled, there is no asymptotic states and that the time scales for transitions into unstable modes are of the order $(2\s)^{-1}$ (see Eq. (\ref{a-number})).
\begin{figure}[t!]
    \includegraphics[width=0.5\textwidth]{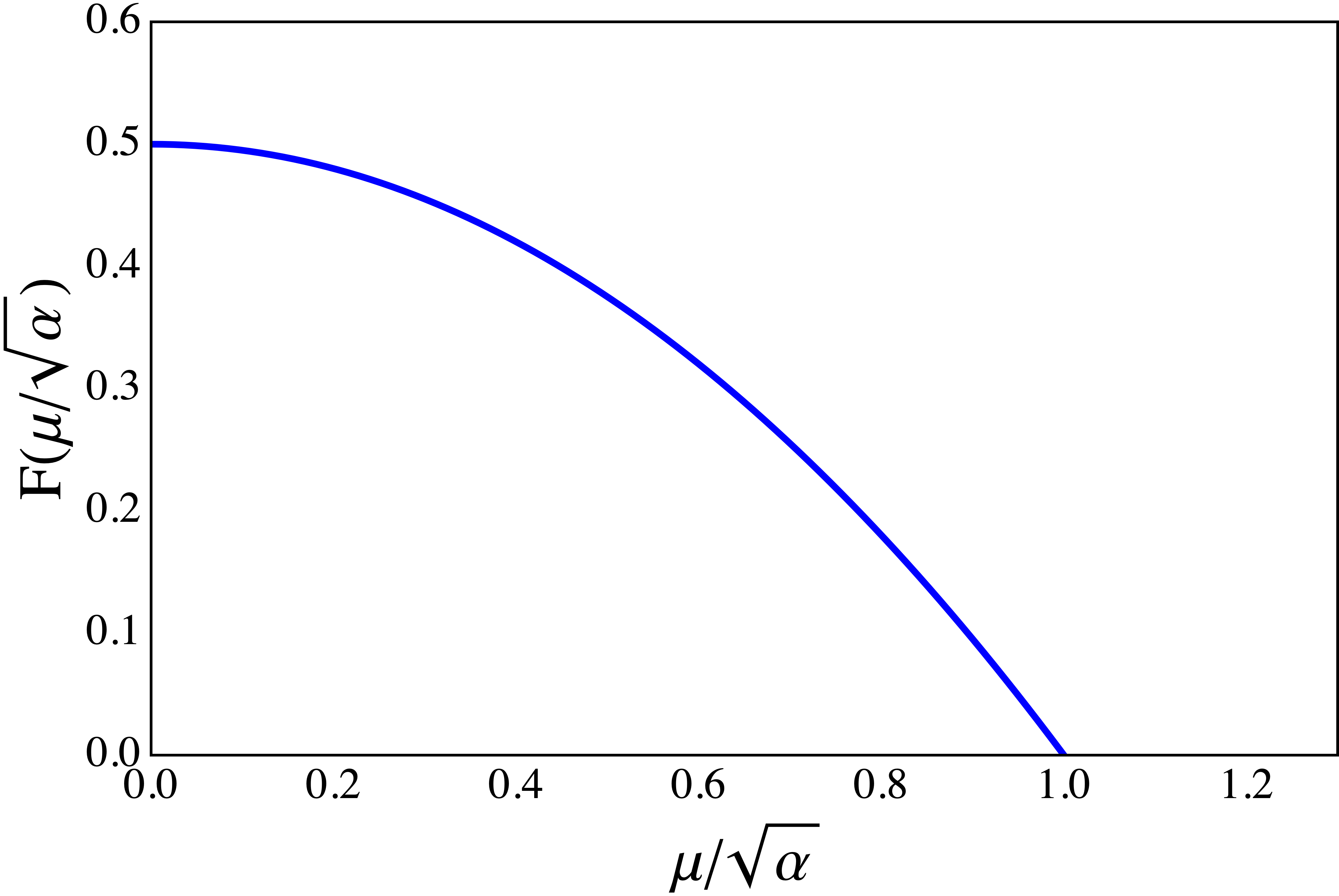}
   \caption{Plot of $F$ as a function of $\mu/\sqrt{\alpha}$. See the text for details.}
    \label{fig4}
\end{figure}

Without saying anything about $T$ we have 
\be
S_{fi}=2g{(2\pi)^3\delta^3(\k+\vec q-\p)\over\sqrt{2\omega_{\p}V}\sqrt{2\Omega_{\k}V}\sqrt{2\Omega_{\vec q}V}}\,e^{i\eps T/2}\,{\sin(\eps T/2)\over\eps/2} \enspace. \label{Smar3}
\ee
Replacing Eq. (\ref{Smar3}) into Eq. (\ref{Gamma1}) and applying an extra factor $1/2$ coming from the fact the
final states are identical particles, we get
\be
\Gamma_{\phi\rightarrow2\chi}={g^2\over2\omega_{\p}}\,\int{d^3k\over(2\pi)^3}(\Omega_{\k}\Omega_{\p-\k})^{-1}{\sin(\eps T)\over\eps} \enspace. \label{Gamma-phi-chi}
\ee

Let us first consider the scenario where $\rho_\phi\rightarrow0$. In this case the system transits to a free of interaction state after the decay, so we can
take $T\rightarrow\infty$. In this limit $\sin(\eps T)/\eps\rightarrow\pi\delta(\eps)$, then the usual energy-momentum conservation relation,
$\eps=\Omega_{\k}+\Omega_{\p-\k}-\omega_{\p}=0$, involving asymptotic initial and final states, holds. For case (1) we, of course, obtain the usual formula
\be
\Gamma_{\phi\rightarrow2\chi}={g^2\over8\pi\mp} \enspace. \label{Gamma-phi-chi0}
\ee
For case (2) we get $\Gamma_{\phi\rightarrow2\chi}=0$ because $\eps$ can never be zero. It is the usual result that prohibits the decay of massless into massive particles due to the lack of final asymptotic states that satisfy energy-momentum conservation.

Now we consider the case where $\rho_\phi$ is big enough to trigger the instabilities discussed before, i.e. it satisfies Eq. (\ref{insta-cond}).
It is clear that there is no asymptotic states for the $\chi$ particles produced from the decay, the evolution of the $\chi$ modes affected by parametric resonance are rather described by Eq. (\ref{sol-a1}).
In parametric resonance the number of the produced particles grows exponentially as Eq. (\ref{a-number}), at least at the beginning,
therefore the transition time of the process is $T=(2\s)^{-1}$. Since the main contributions for the growth are provided by modes that fulfill $\s\gg\eps$,
we can use the approximation $\sin(\eps T)=\sin(\eps/2\s)\approx\eps/2\s$ with great accuracy. Equation (\ref{Gamma-phi-chi}) becomes
\be
\Gamma_{\phi\rightarrow2\chi}={g^2\over2\omega_{\p}}\,\int{d^3k\over(2\pi)^3}{(\Omega_{\k}\Omega_{\p-\k})^{-1}\over\sqrt{4\sig^2-\eps^2}} \enspace. \label{Gamma-phi-chi-high}
\ee

For case (1) we can make $\sig\approx\mp\alpha$ to preserve only terms of order $g^2$. We find
\beq
\Gamma_{\phi\rightarrow2\chi}^{(1)} &=& {g^2\over8\pi^2m_\phi}\int_{-2\mp\alpha}^{2\mp\alpha}{d\epsilon\over\sqrt{4\mp^2\alpha^2-\epsilon^2}}
\\
&=& {g^2\over8\pi\mp} \enspace. \label{Gamma-phi-chi-high-1}
\eeq
As we get the same as Eq. (\ref{Gamma-phi-chi0}), the physical reason of this result should be investigated further.

For case (2) the
integration is not straightforward as in case (1). Assuming $\theta\ll1$, which is true on behalf of previous discussions,  we get
\be
\Gamma_{\phi\rightarrow2\chi}^{(2)}={g^2\over8\pi p}F(\mu/\sqrt{\alpha}) \label{Gamma-phi-chi-high-2}
\ee
where
\be
F(x)={1\over\pi}\int_{\kappa_-(x)}^{\kappa_+(x)}d\kappa\left({\pi\over2}-\arcsin\left({x^2\over2\sqrt{\kappa\,(1-\kappa)}}\right)\right) \enspace. \label{F}
\ee
We recall that $\kappa_{\pm}(\mu/\sqrt{\alpha})$ are defined in Eq. (\ref{kpm}). In terms of the quantity $\omega_{\p}$, which is $\mp$ for case (1)
and $p$ for case (2), the ratio between both decay rates is simply
\be
{\Gamma_{\phi\rightarrow2\chi}^{(2)}\over\Gamma_{\phi\rightarrow2\chi}^{(1)}}=F(\mu/\sqrt{\alpha}) \enspace. \label{ratio}
\ee
A plot of $F(\mu/\sqrt{\alpha})$ is shown in Fig. \ref{fig4}. As we already discussed,  we can see that for case (2) the decay is clearly possible if $\mu<\sqrt{\alpha}$,
but it becomes impossible for $\mu\geq\sqrt{\alpha}$. For $\mu\ll\sqrt{\alpha}$ the decay rate approaches its maximum which is a half of the decay rate for case (1). 

In our toy model we can also connect the $g^2$ dependence of the earliest spontaneous decay rate and the rate proportional to $g$ at which $\chi$ particles are produced by parametric resonance.
As for the axion-photon discussion in Sec. \ref{axioncase}, we take the Boltzmann equation $\dot n_\phi=-\Gamma_{\phi\rightarrow2\chi}(1+2f_{\chi})n_\phi$, where
$f_{\chi}\sim(2\pi)^3n_\chi/d^3k$ and $\Gamma_{\phi\rightarrow2\chi}\sim g^2/(8\pi\omega_{\p})$ for both case (1) and case (2).  It is not difficult to find that also for both cases
$d^3k\sim\pi g\sqrt{2\rho_\phi}$. Combining all of this and the fact that $\dot n_\chi=-2\dot n_\phi$, we find from the Boltzmann equation that $\dot n_\chi\sim{g\sqrt{2\rho_\phi}\over\omega_{\p}^2}n_\chi$,
which is consistent with Eq. (\ref{a-number}) for case (1) as well as for case (2).

\section{Depletion of the massless beam}

To finish we will briefly discuss the conditions that allow a significant depletion of the $\phi$ field. We already said that condition Eq. (\ref{insta-cond})
triggers the instability, however it is a necessary but not sufficient condition for a substantial decay of the beam. The missing condition is
that the time $t_\text{res}$ during which the instability takes place is enough to reach the exponential regime. It requires $\alpha\,\omega_{\p}\,t_\text{res}>1$.
We need the produced particles
to stay inside the beam extension for the parametric resonance
to develop, therefore $t_\text{res}$ is the time that $\chi$ particles take to leave the beam region. As the $\chi$ particles have momenta pointing mainly to the same direction as the beam propagates,
$t_\text{res}$ is given roughly by $d/(1-v_\chi)$, where $v_\chi$
is the velocity of the produced particles and $d$ the beam length. This velocity is approximately $1-\mu^2/2$, which leads to the following depletion condition
\be
d>{\mu^2\over2\alpha p} \enspace. \label{dep-cond}
\ee
Notice that, given the resonance condition $\mu^2<\alpha$, Eq. (\ref{dep-cond}) is satisfied necessarily by taking $d>1/(2p)$.

\section{Conclusion}

In this article we have taken a simple scalar toy model, with interaction term $g\phi\chi^2$, to study the process $\phi\rightarrow2\chi$ in
the particular case of massless $\phi$ and massive $\chi$. Although the one particle analysis prohibits the decay by the requirement of energy-momentum
conservation of asymptotic states, we found that when $\phi$ is highly occupied, nonperturbative effects (parametric resonance) allow the decay as long as the energy density
of the decaying particle beam exceeds the threshold defined in Eq. (\ref{insta-cond}). We demonstrated that energy-momentum is indeed conserved during the particle production and, finally,
we got an analytical expression for the spontaneous decay rate at the earliest time. This result suggests a deeper inspection of nonperturbative effects in particle processes, especially the ones that are relevant in
cosmology, astrophysics and collider physics.


\acknowledgments

We would like to thank Ilya Bakhmatov, Crist\'obal Corral, Jorge Gamboa, Benjamin Koch, Dmitry Levkov, Ernesto Matute, Fernando M\'endez,
Guillermo Palma, Dmitry Ponomarev, Pierre Sikivie and Elisa Todarello for useful discussions as well as Joerg Jaeckel and Thomas Schwetz for
their feedback and suggestions. We want to give a special thanks to Paola Arias and Diego Vargas for that discussion where this idea arose
and Crist\'obal Garrido for his great support here in Moscow.

\appendix

\section{Technical details for solutions in the rotating wave approximation} \label{RWA}

Let us study the resonant solutions of Eq. (\ref{eqA1}) analytically using the rotating wave approximation (RWA). To do so we first write $\chi_{\k}(t)=a_{\k}(t)\,e^{-i\Omega_{\k}\,t}$
where $a_{\k}$ varies slowly respect to $\chi_{\k}$. Neglecting second derivatives of $a_{\k}$, Eq. (\ref{eqA1}) can be written as
\beq
\partial_ta_{\k} &=& \partial_ta_{-\k}^\dagger\,e^{2i\Omega_{\k}\,t} -i\sigma_{\k}\left(a_{\k-\p}\,e^{i\epsilon_{\k}^{(1)}t}+a_{\p-\k}^\dagger\,e^{i\epsilon_{\k}^{(2)}t}\right) \nonumber
\\
& & -i\sigma_{\k+\p}\left(a_{\k+\p}\,e^{i\epsilon_{\k}^{(3)}t}+a_{-\p-\k}^\dagger\,e^{i\epsilon_{\k}^{(4)}t}\right), \label{eqa1s}
\eeq
where
\be
\sig=g\sqrt{\rho_\phi/2\over\omega_{\p}^2\,\Omega_{\p-\k}\,\Omega_{\k}} \label{sigmaks}
\ee
and
\beq
\epsilon_{\k}^{(1)} &=& \Omega_{\k}-\Omega_{\k-\p}-\omega_{\p} \label{eps1s}
\\
\epsilon_{\k}^{(2)} &=& \Omega_{\k}+\Omega_{\p-\k}-\omega_{\p} \label{eps2s}
\\
\epsilon_{\k}^{(3)} &=& \Omega_{\k}-\Omega_{\k+\p}+\omega_{\p} \label{eps3s}
\\
\epsilon_{\k}^{(4)} &=& \Omega_{\k}+\Omega_{-\k-\p}+\omega_{\p} \enspace. \label{eps4s}
\eeq
The RWA allows us to only keep terms that oscillate slowly respect to $\Omega_{\k}$. Therefore, we can immediately neglect the first term
of the RHS of Eq. (\ref{eqa1s}) of this document as well as the term containing $\epsilon_{\vec k}^{(4)}$. The recognition of the other terms that can be neglected will depend on $\k$.
The $\epsilon_{\k}^{(i)}$ defined in Eqs. (\ref{eps1s})-(\ref{eps3s}) account for the efficiency of some particular processes. This association is listed as follows
\beq
\epsilon_{\k}^{(1)}:\ \ \ \ \ \ \ \  &  \phi_{\p}+\chi_{\k-\p}\,  \leftrightarrow\,   \chi_{\k}  \label{process1s}
\\
\epsilon_{\k}^{(2)}:\ \ \ \ \ \ \ \ &  \phi_{\p}\,  \leftrightarrow\,   \chi_{\k}+\chi_{\p-\k}  \label{process2s}
\\
\epsilon_{\k}^{(3)}:\ \ \ \ \ \ \ \ &  \phi_{\p}+ \chi_{\k}\,  \leftrightarrow\,   \chi_{\k+\p} \enspace. \label{process3s}
\eeq
When one of them becomes smaller than some threshold, the associated process is excited. As in our initial condition there is no
$\chi$ particles, at the very beginning only the process $\phi_{\p}\rightarrow\chi_{\k}+\chi_{\p-\k}$ can be possible, so for early times Eq. (\ref{eqa1s}) reduces to
\be
\dot a_{\k}=-i\sig\,a_{\p-\k}^\dagger\,e^{i\epsilon_{\k}^{(2)}\,t} \enspace. \label{eqa2s}
\ee

\section{Further processes after first decays} \label{processes}

This articler is mainly based on the solution of Eq. (\ref{eqa2s}), which gives a complete picture of the system at early times, where $\chi$ is not significantly populated. In this time scale, only the process $\phi_{\p}\rightarrow\chi_{\k}+\chi_{\p-\k}$ can be excited. Now we are going
to discuss what happens after the modes $\k$ and $\p-\k$ are appreciably occupied. Even at this times, we are still assuming that $\phi$ is not depleted substantially, so the linearizaion of the equations of motion still holds.
\begin{figure}[t!]
    \includegraphics[width=0.5\textwidth]{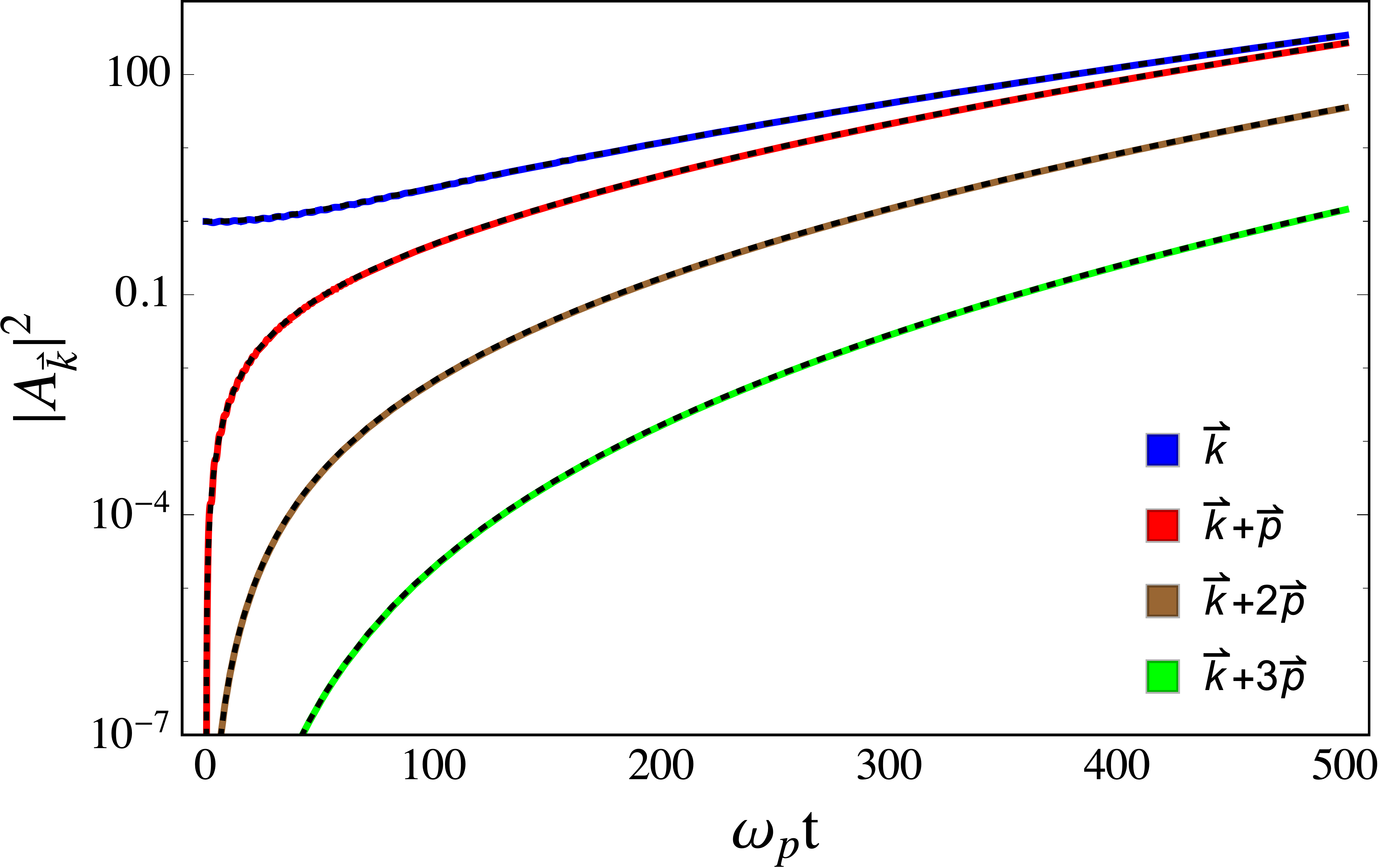}
   \caption{Plots of the amplitude squared $|A_{\k}|^2$ as a function of time for the modes $\k$, $\k+\p$, $\k+2\p$ and $\k+3\p$. Dotted lines correspond to numerical solutions using the rotating wave approximation. See the text for more details.}
    \label{fig}
\end{figure}

For case (1) (massive $\phi$ and massless $\chi$) none of the other $\epsilon_{\vec k}^{(i)}$ defined in Eq. (\ref{eps1s}) and Eq. (\ref{eps3s}) become small enough to excite the corresponding process,
then $\phi$ continue decaying into momentum modes $\k$ and $\p-\k$ of $\chi$ particles and Eqs. (\ref{sol-a1}) and (\ref{a-number}) are still valid. Their validity holds until back-reactions are activated, i.e., when the inverse process
$2\chi\rightarrow\phi$ starts to be important. Of course, it happens when $\phi$ is significantly depleted.

For case (2) $\epsilon_{\k}^{(1)}$ is big, of the order of $\Omega_{\k}$, but $\epsilon_{\k}^{(3)}$
is as small as $\epsilon_{\k}^{(2)}$. It means that if the process $\phi_{\p}\rightarrow\chi_{\k}+\chi_{\p-\k}$ is efficient, once $\chi_{\k}$ and $\chi_{\p-\k}$ are abundant, the processes $\chi_{\k}+\phi_{\p}\rightarrow\chi_{\k+\p}$ and $\chi_{\p-\k}+\phi_{\p}\rightarrow\chi_{2\p-\k}$ will become efficient too. Following the same reasoning, a cascade is produced and every process of the form $\chi_{\k+n\p}+\phi_{\p}\rightarrow\chi_{\k+(n+1)\p}$ and $\chi_{n\p-\k}+\phi_{\p}\rightarrow\chi_{(n+1)\p-\k}$ ($n=0, 1, 2,...$) will be eventually important after some period of time. To prove this last claim, we compute the classical version of Eq. (\ref{eqA1}) numerically assuming that only the modes $\k$ and $\p-\k$ are initially occupied. Figure \ref{fig} shows $|A_{\k}|^2$ as a function of time for the modes $\k$, $\k+\p$, $\k+2\p$ and $\k+3\p$. We do not plot the modes $\p-\k$, $2\p-\k$, $3\p-\k$ and $4\p-\k$ because, according to the initial conditions, the time evolution of their square amplitudes are identical as the ones for $\k$, $\k+\p$, $\k+2\p$ and $\k+3\p$, respectively. We evaluate at $\k=\omega_{\p}/2$ and we assume that at the beginning the modes $\k$ and $\p-\k$ have initial conditions $|A_{\k}(0)|^2=|A_{\p-\k}(0)|^2=1$. All the other modes have a null initial amplitude. We see that the squared amplitudes of modes different from $\k$ and $\p-\k$ begin to grow gradually. In Fig. \ref{fig} we also compute numerically in the rotating wave approximation (see dotted lines). We basically take Eq. (\ref{eqa1s}) and solve the system neglecting fast oscillating terms. It is clear that this approximation perfectly agrees with the full solutions of the second order differential equations. 
\newline
\newline

\section{Energy expectation value for the produced field} \label{energy}

Inserting Eq. (\ref{chi1}) of the paper, with $\chi_{\k}=a_{\k}\,e^{-i\Omega_{\k}t}$, into the hamiltonian defined in Eq. (\ref{Hchi}), we obtain
\beq
H_\chi &=& {1\over2}\int{d^3k\over(2\pi)^3}\left(\Omega_{\k}(a_{\k}\,a_{\k}^\dagger+a_{\k}^\dagger\,a_{\k})\right. \nonumber
\\
& & \left.+{i\over2}(\dot a_{\k}\,a_{\k}^\dagger+a_{\k}^\dagger\,\dot a_{\k}-a_{\k}\,\dot a_{\k}^\dagger-\dot a_{\k}^\dagger \,a_{\k})\right) \enspace. \label{H1s}
\eeq
To get Eq. (\ref{H1s}) we were consistent with the RWA, therefore we have neglected fast oscillating terms as well as terms of order $\dot a^2$, $(\dot a^\dagger)^2$, $\dot a\dot a^\dagger$ and $\dot a^\dagger\dot a$. Now we calculate the expectations values of each term in the integrand using the solution Eq. (\ref{sol-a1}).
Ignoring the contribution of the vacuum energy, we have 
\be
\left<0\right|a_{\k}\,a_{\k}^\dagger+a_{\k}^\dagger\,a_{\k}\left|0\right>=2Vf_{\chi,\k} \enspace.
\ee
For the other terms we get
\beq
\left<0\right|\dot a_{\k}\,a_{\k}^\dagger\left|0\right> &=& V{\sig^2\over\s}S_{\k}\left(C_{\k}+i{\eps\over2\s}S_{\k}\right)
\\
\left<0\right|a_{\k}\,\dot a_{\k}^\dagger\left|0\right>  &=& V{\sig^2\over\s}S_{\k}\left(C_{\k}-i{\eps\over2\s}S_{\k}\right)
\\
\left<0\right|a_{\k}^\dagger\,\dot a_{\k}\left|0\right> &=& \left<0\right|\dot a_{\k}\,a_{\k}^\dagger\left|0\right>
\\
\left<0\right|\dot a_{\k}^\dagger \,a_{\k}\left|0\right>  &=& \left<0\right|a_{\k}\,\dot a_{\k}^\dagger\left|0\right> \enspace,
\eeq
where $C_{\k}=\cosh(\s t)$ and $S_{\k}=\sinh(\s t)$. With these results, we find
\be
\left<E_\chi\right>=\left<0\right|H_\chi\left|0\right>=V\int {d^3k\over(2\pi)^3}(\Omega_{\k}-\e/2)f_{\chi,\k} \enspace. \label{Echi1s}
\ee


\bibliographystyle{JHEP}
\bibliography{references.bib}

\end{document}